\documentclass[pre,aps,twocolumn,amsmath,amssymb]{revtex4-1}

\pdfoutput=1
\usepackage{hyperref}
\usepackage{graphicx}
\usepackage[usenames]{color}
\newcommand{\new}[1]{\textcolor{black}{#1}}

\def\f#1{Fig.~\ref{#1}}

\def\c#1{~\cite{#1}}

\def\beq{\begin{equation}}
\def\eeq{\end{equation}}
\def\bea{\begin{eqnarray}}
\def\eea{\end{eqnarray}}

\def\kt{k_{\rm B}T}

\begin{document}

\title{Minimal positive design for self-assembly of the Archimedean tilings}
\author{Stephen Whitelam}\email{{\tt swhitelam@lbl.gov}}
\address{Molecular Foundry, Lawrence Berkeley National Laboratory, 1 Cyclotron Road, Berkeley, CA 94720, USA}

\begin{abstract}
A challenge of molecular self-assembly is to understand how to design particles that self-assemble into a desired structure and not any of a potentially large number of undesired structures. Here we use simulation to show that a strategy of {\em minimal positive design} allows the self-assembly of networks equivalent to the 8 semiregular Archimedean tilings of the plane, structures not previously realized in simulation. This strategy consists of identifying the fewest distinct types of interparticle interaction that appear in the desired structure, and does not require enumeration of the many possible undesired structures. The resulting particles, which self-assemble into the desired networks, possess DNA-like selectivity of their interactions. Assembly of certain molecular networks may therefore require such selectivity. 
\end{abstract}

\maketitle

Molecular self-assembly is the spontaneous organization of molecules or other small particles into structures\new{\c{whitesides1991molecular,biancaniello2005colloidal,park2008dna,nykypanchuk2008dna, chen2011directed,fan2011dna, ke2012three,feng2013specificity,tian2016lattice,liu2016self}}. Despite many successes in the laboratory we lack complete understanding of how to design particles that will self-assemble into a desired structure: sometimes the outcome of self-assembly is an undesired structure, which might be metastable or kinetically trapped\c{whitelam2015statistical}. Simulation can help us understand how particle design influences the process and outcome of self-assembly\c{doye2004inhibition,hagan2006dynamic,doye2007condensed,rapaport2010modeling,reinhardt2014numerical,murugan2015undesired}. \new{Ideas derived from such studies include the notion of {\em positive} and {\em negative design}, used by Doye, Louis and Vendruscolo\c{doye2004inhibition} to describe the design of particles to promote a desired structure or to suppress undesired structures. We show here that a particular type of positive design, which we shall call {\em minimal positive design}, allows the self-assembly of network structures equivalent to the Archimedean tilings of the plane.} The concept is simple to implement: we drew pictures of the desired structures and labeled the interparticle interactions that arise, and then (on the computer) made particles with those interactions {\em and no others}. Those particles self-assembled, under a simple cooling protocol, into the desired structures, which had not previously been realized in simulation\c{antlanger2011stability}. \new{The field of DNA nanotechnology makes widespread use of the principle of chemical selectivity\c{seeman1998dna,jones2010dna,tkachenko2011theory,pfeifer2016nano}, allowing, for instance, self-assembly of structures in which particles are of many distinct types\c{ke2012three,halverson2013dna,zeravcic2014self}. Here we show that chemically selective interactions are also necessary to assemble certain complex single-component structures, which a priori do not appear to require this capability. By formalizing the idea of minimal positive design we hope to provide a way of thinking about chemical selectivity and its role in self-assembly.}
\begin{figure*}[t!] 
   \centering
\includegraphics[width=0.9\linewidth]{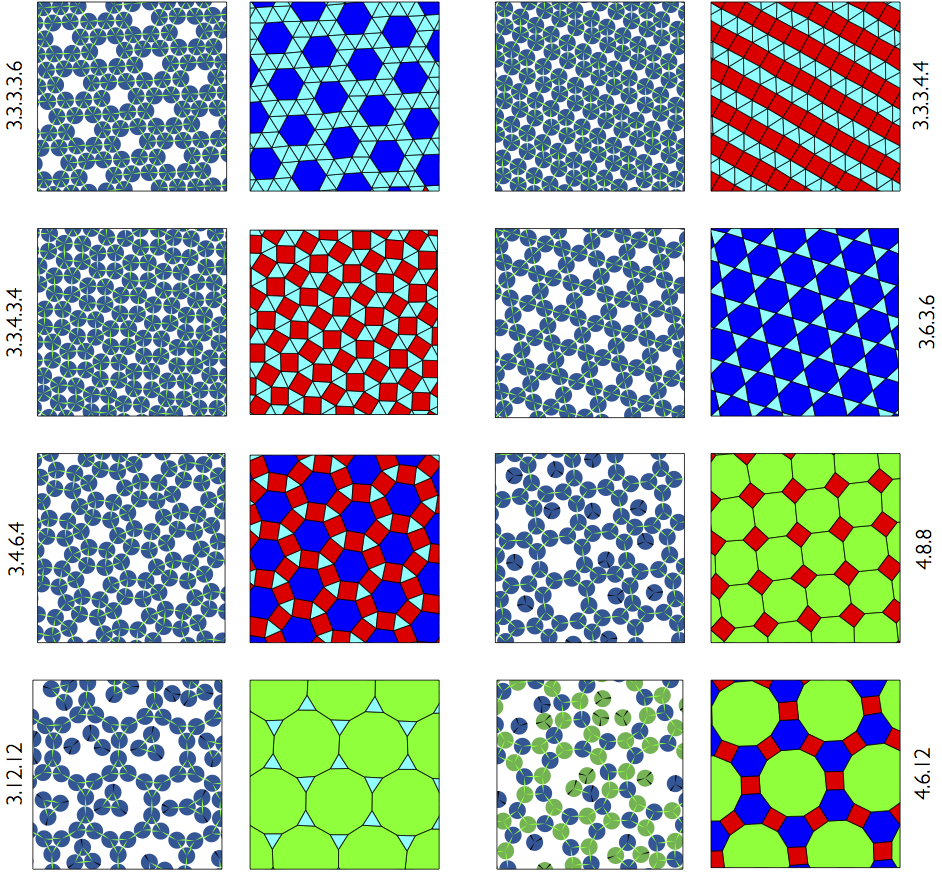} 
   \caption{\label{fig1} Networks of patchy particles (left images), and their tiling representations (right images) obtained by joining the centers of interacting particles. These networks, equivalent to the 8 semiregular Archimedean tilings, self-assembled from particles of the geometry and attractive patch-patch interactions given in Table \ref{tab1}.}
\end{figure*}

We illustrate the concept of minimal positive design by considering the self-assembly of networks equivalent to tilings of the plane\c{grunbaum1977tilings, kepler1938harmonice}. Such networks are complex geometrically, and so pose a challenge for design, and their experimental realizations, via the self-assembly of real molecules at surfaces, have useful properties\c{ecija2013five,elemans2009molecular,bartels2010tailoring,whitelam2014common}. The simplest tilings of the plane are the 3 Platonic tilings, which consist of regular triangles, squares or hexagons. The networks equivalent to these tilings can be self-assembled from {\em patchy particles}, model molecules with anisotropic pairwise interactions\c{zhang2004self,doye2007controlling,bianchi2008theoretical,duguet2016patchy}, that possess regular sixfold, fourfold or threefold rotational symmetry, respectively\c{doppelbauer2010self,antlanger2011stability,doye2007controlling,whitelam2015emergent}. Evident here is the concept of positive design: particles have the same rotational symmetry as the vertices of the network equivalent to the edges of the tiling. In addition, the formation of alternative structures is avoided during self-assembly, and the desired networks spontaneously appear.

Networks equivalent to the 8 Archimedean tilings of the plane pose a sterner challenge for design. These tilings, sections of which are shown in \f{fig1}, are composed of two or three polygon types, but only one type of vertex\c{grunbaum1977tilings}\footnote{With one exception, as we shall discuss.}. In principle, therefore, they can be made from a single type of particle~\footnote{Self-assembly of the Archimedean tilings has been done on the computer using particles that represent the {\em tiles} of the tilings\c{antlanger2011stability,millan2014self}, or by using mixtures of particles\c{salgado2015non}; in these cases one needs in general more than one type of particle per tiling.}. However, Ref.\c{antlanger2011stability} showed that patchy particles whose interaction geometries are those of the vertices of these tilings are not in general thermodynamically stable in network form, tending instead to form more compact undesired structures. Consistent with this observation we found that such particles do not in general self-assemble into networks equivalent to the Archimedean tilings. 
\begin{figure*}[] 
   \centering
\includegraphics[width=0.8\linewidth]{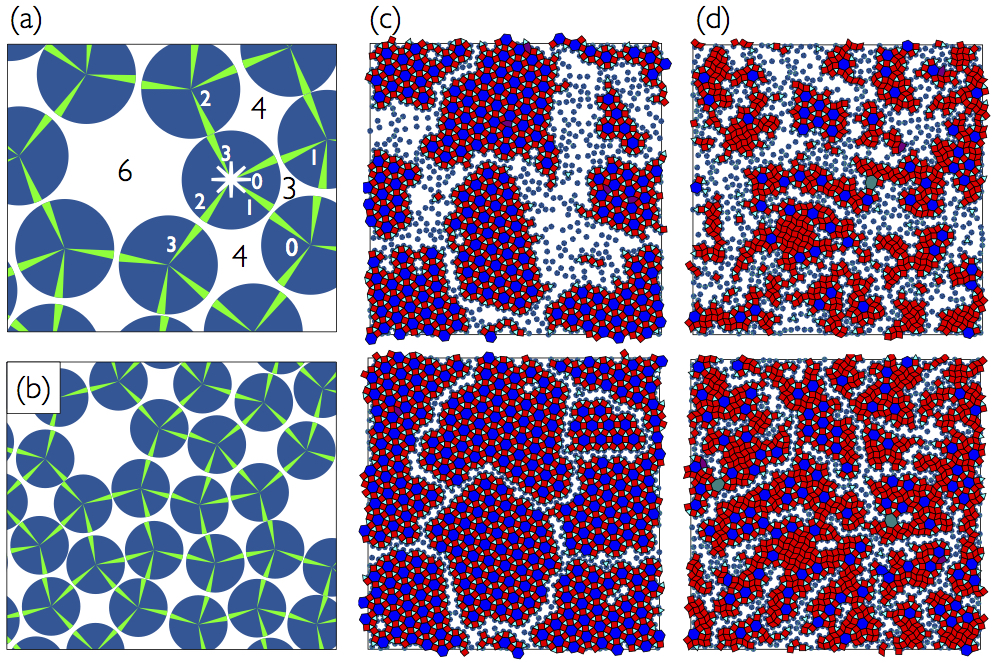} 
   \caption{\label{fig2} (a) The starred particle acts as a vertex of the network equivalent to the 3.4.6.4 tiling (the surrounding tile types are labeled in black). The only patch-to-patch contacts made are 0-to-1 and 2-to-3 (patch identities are marked in white). (b) If other patch-to-patch contacts are permitted then the particle can form other networks. (c) Simulations done in the presence of only the necessary interactions result in self-assembly of the 3.4.6.4 network (we draw convex polygons on top of the network). (d)  Simulations done when all patch-patch interactions are attractive result in undesired structures.}
\end{figure*}

The concept of minimal positive design can be used to overcome this problem. In \f{fig2}(a) we show a section of a network equivalent (in an averaged sense) to the $3.4.6.4$ Archimedean tiling. Archimedean tilings are designated $a_0.a_1.\cdots.a_{K-1}$, meaning that as we move in a circle around any vertex we encounter a regular $a_0$-gon, followed by a regular $a_1$-gon, and so on, ending with a regular $a_{K-1}$-gon\c{grunbaum1977tilings}. The network shown is made from a 4-patch particle with an angle of $\theta(3) = 60^\circ$ between the bisectors of patches 0 and 1, an angle of $\theta(4) = 90^\circ$ between the bisectors of patches 1 and 2, an angle of $\theta(6) = 120^\circ$ between the bisectors of patches 2 and 3, and (therefore) an angle of $\theta(4) = 90^\circ$ between the bisectors of patches 3 and 0; here $\theta(n) \equiv (n-2)\pi/n$ is the internal angle of a regular $n$-gon. We shall call this a $3.4.6.4$ particle. In general, as noted in\c{antlanger2011stability}, a vertex of an Archimedean tiling $a_0.a_1.\cdots.a_{K-1}$ can be made from a patchy particle of the same destination, i.e. one with $K$ patches whose bisectors are separated by angles $\theta(a_0),\theta(a_1),\dots,\theta(a_{K-1})$ (we label patches by integers $0,1,\dots,K-1$ so that $\theta(a_i)$ is the angle between the bisectors of patches $i$ and $i+1$, with patch $K$ being also patch 0).

If all patches of the particle shown in \f{fig2}(a) attract all other patches then the particle can participate in undesired networks like that shown in \f{fig2}(b), and in a large number of disordered networks. Such networks compete thermodynamically with the target structure\c{antlanger2011stability}, or act as traps that kinetically arrest assembly. However, looking again at \f{fig2}(a) we see that not all patch-patch interactions appear in the $3.4.6.4$ tiling: only the pairs 0:1 and 2:3 do so. If we construct a patchy particle whose patches are chemically selective in that way, allowing only attractions between patch 0 and 1 and between patch 2 and 3, then, as shown in \f{fig3}(c), it self-assembles into the desired network. This chemical selectivity is the {\em minimal} element in the strategy of minimal positive design. Absent this selectivity the self-assembly process results in the undesired structure shown in \f{fig2}(d).

\begin{figure*}[t!] 
   \centering
\includegraphics[width=0.8\linewidth]{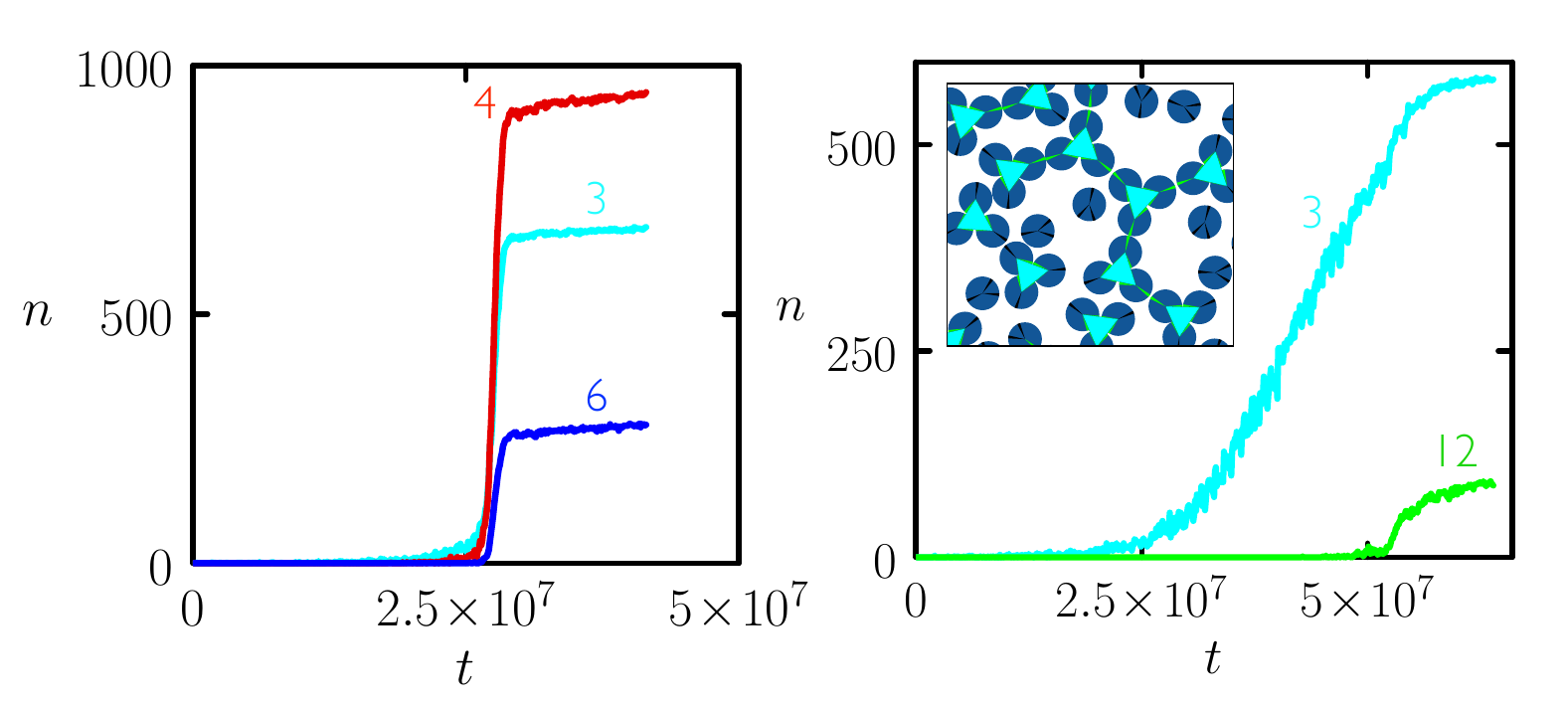} 
   \caption{\label{fig3} Number of convex polygons ($n$), of type indicated, as a function of Monte Carlo step number ($t$) during assembly of the 3.4.6.4 tiling (left) and the 3.12.12 tiling (right). The 3.4.6.4 tiling assembles via nucleation and growth of the unit cell; the 3.12.12 tiling results instead from the coarsening of a network of triangles, a piece of which is shown in the inset.}
\end{figure*}
\begin{table}[h]
\caption{Interactions for the minimal positive design of the Archimedean tilings.}
\begin{center}
\begin{tabular}{c|c}
{\rm particles} & {\rm complementary patches} \\
\hline 
3.3.3.3.6 & 0:4, 1:1, 2:3 \\
3.3.3.4.4 & 0:3, 1:1, 2:2, 4:4 \\
3.3.4.3.4 & 0:4, 1:1, 2:3 \\
3.6.3.6 & 0:1, 2:3 \\
3.4.6.4 & 0:1, 2:3 \\
4.8.8 & 0:1, 2:2 \\
3.12.12 & 0:1, 2:2 \\
4.6.12 {\rm and} 4.12.6 & 0:1$'$, 2:2$'$
\end{tabular} \end{center} \label{tab1} \end{table}
The simulation model used to make these figures is similar to that used in Ref.\c{whitelam2015emergent}. The particles in question are hard discs of diameter $a$. Particles attract each other via patches of opening angle $10^\circ$. Particle interactions are square-well in both angle\c{kern2003fluid} and range, and they are chemically selective: particles possess a pairwise attraction of energy $-\epsilon \, \kt$ if 1) two disc centers lie within a distance $11a/10$; if 2) the line joining those two discs cuts through one patch on each disc; and if 3) the patch-patch pairs are those shown in Table~\ref{tab1}. Patches engaged in this manner are shown green in figures, and those not engaged are shown black. For illustrative purposes we sometimes draw the convex polygons that result from joining the centers of bound discs. The simulation protocol we used is designed to mimic the deposition and assembly of real molecules on surfaces. We work in the $\mu VT$ ensemble. We start with an empty substrate and allow particles to appear and disappear on it using grand-canonical Monte Carlo moves\c{frenkel1996understanding}. We allow particles to move on the substrate using the virtual-move Monte Carlo algorithm described in the appendix of Ref.\c{whitelam2009role} (\new{we checked that the results of several simulations were qualitatively unchanged upon using also a Metropolis Monte Carlo dynamics}). We started the simulation with a small value of $\epsilon$ ($\approx 1$) and a chemical potential chosen so that the substrate is sparsely occupied with particles, and we `cooled' the system slowly by increasing $\epsilon$ by a value of $\approx 0.1$ every million Monte Carlo steps (\new{this is a convenient way of finding the bond energy at which assembly can proceed}). Eventually, we observe the self-assembly of a network structure.

Inspection of the Archimedean tilings\c{grunbaum1977tilings} reveals that the corresponding networks can be built from the particle geometries and interactions listed in Table~\ref{tab1}. We verified that particles of this nature self-assemble in simulations into networks equivalent to the Archimedean tilings. Sections of these networks are shown in \f{fig1}; simulation boxes from which these sections are taken are shown in Fig. S1. Networks possess some disordered regions, such as grain boundaries, but otherwise cover large portions of the substrate. Networks are porous, and in some of the larger pores we observe the assembly of smaller structures. Pictures of the networks alone are shown in Fig. S2. Most of these networks emerge via nucleation and growth of the ordered unit cell. One that does not is the 3.12.12 network, which self-assembles hierarchically: a network of triangles forms and subsequently rearranges to complete the 12-gons; see \f{fig3}. 

One point of note, mentioned in Ref.\c{antlanger2011stability}, is that the 4.6.12 tiling must be made from two types of vertex (or particle), 4.6.12 and 4.12.6. These vertices are usually considered equivalent in the mathematical literature\c{grunbaum1977tilings}, but the corresponding particles are distinct objects. Thus the 4.6.12 tiling is unlike the other 7 Archimedean tilings in that it requires at least two types of particle for its construction. The primes in the last row of Table~\ref{tab1} indicate that the required attractive interactions are between the two types of particle. We used an equimolar mixture of the two particle types.

In Fig. S3 we show the results of self-assembly done without the constraint of chemical selectivity. Sections of the 3.3.3.4.4 and 3.3.4.3.4 tilings appear. The 3.6.3.6 tiling appears, albeit in competition with a rhombic polymorph (which would be favored at finite pressure\c{antlanger2011stability}); see Fig. S4. In general, however, we do not observe large-scale self-assembly of networks equivalent to the Archimedean tilings.

We have shown that the strategy of minimal positive design -- geometry plus chemical selectivity -- succeeds in self-assembling networks equivalent to the Archimedean tilings where positive design alone -- geometry -- fails. \new{Self-assembly using minimal positive design is still vulnerable, like other examples of self-assembly, to kinetic traps that can thwart assembly\c{hagan2006dynamic,doye2007controlling,whitelam2015statistical}, particularly when assembly intermediates are large and floppy, because they can be prevented from closing by adjacent structures. Such traps can be exacerbated e.g. by rapid cooling, causing assembly to fail. Importantly, though, there are conditions for which assembly can succeed, which is not true here for the case of positive design alone.} The strategy of minimal positive design used here is simple to implement. It consists only of the enumeration of the interactions that occur in desired structures, and does not involve assessment of the thermodynamic stability of the desired structures, nor explicit consideration of possible competing structures (\new{many of which are disordered and would be difficult to enumerate}).  \new{Although not all structures can be thought of in the terms considered here -- in quasicrystals, for instance, one cannot uniquely enumerate the interactions made by every particle in the structure\c{reinhardt2016self} -- many can}. \new{Moreover, while assembly of networks equivalent to the simplest (Platonic) tilings of the plane does not require chemical specificity of interactions, assembly of networks equivalent to the next-simplest (Archimedean) tilings does. It seems likely, therefore, that several areas of molecular self-assembly -- involving e.g. molecules at surfaces -- could benefit from an interaction principle that is widely seen in proteins\c{jones1996principles,fusco2014characterizing}, and used in DNA nanotechnology to create complex structures\c{seeman1998dna,pfeifer2016nano,zhang2009symmetry}.}

\begin{acknowledgments}
{\em For A\"ida, Marc, and Aria.} I thank Barbara Sacc{\`a} for valuable discussions. This work was done at the Molecular Foundry, Lawrence Berkeley National Laboratory, and was supported by the Office of Science, Office of Basic Energy Sciences, of the U.S. Department of Energy under Contract No. DE-AC02--05CH11231
\end{acknowledgments}


%

\onecolumngrid
\clearpage

\renewcommand{\theequation}{S\arabic{equation}}
\renewcommand{\thefigure}{S\arabic{figure}}
\renewcommand{\thesection}{S\arabic{section}}

\setcounter{equation}{0}
\setcounter{section}{0}
\setcounter{figure}{0}

\setlength{\parskip}{0.25cm}%
\setlength{\parindent}{0pt}%

\section{Supplementary figures}
\label{supp1}

\begin{figure*}[h!] 
   \centering
\includegraphics[width=\linewidth]{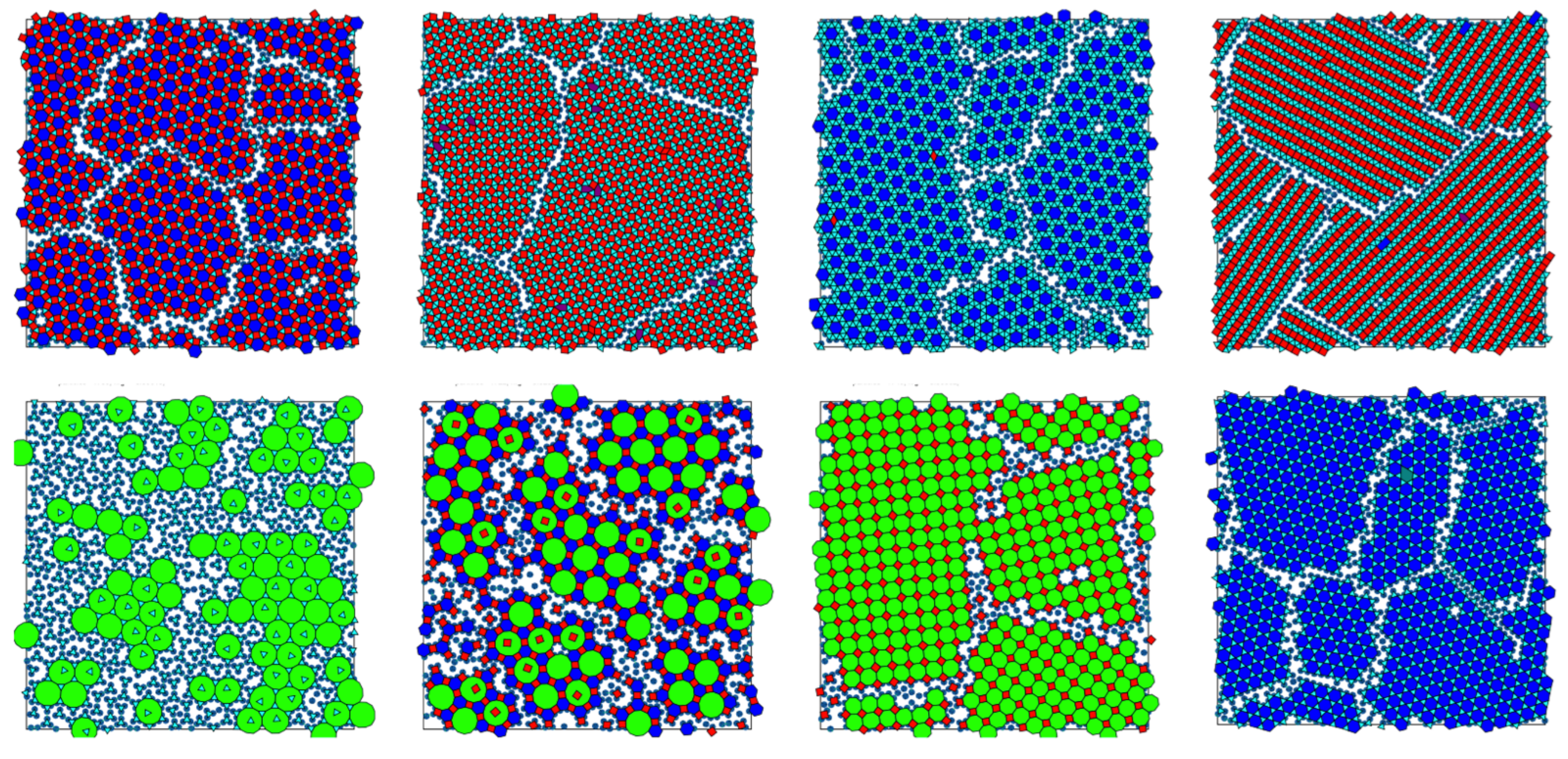} 
   \caption{\label{figs1} Simulation boxes from which the snapshots in \f{fig1} are taken.}
\end{figure*}

\break

\begin{figure*}[h!] 
   \centering
\includegraphics[width=\linewidth]{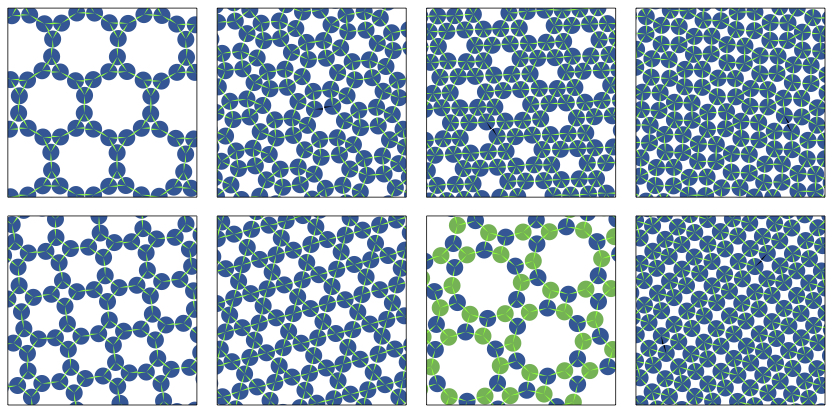} 
   \caption{\label{figs2} The networks of \f{fig1} shown, for clarity, without the particles inside the pores.}
\end{figure*}

\break

\begin{figure*}[h!] 
   \centering
\includegraphics[width=\linewidth]{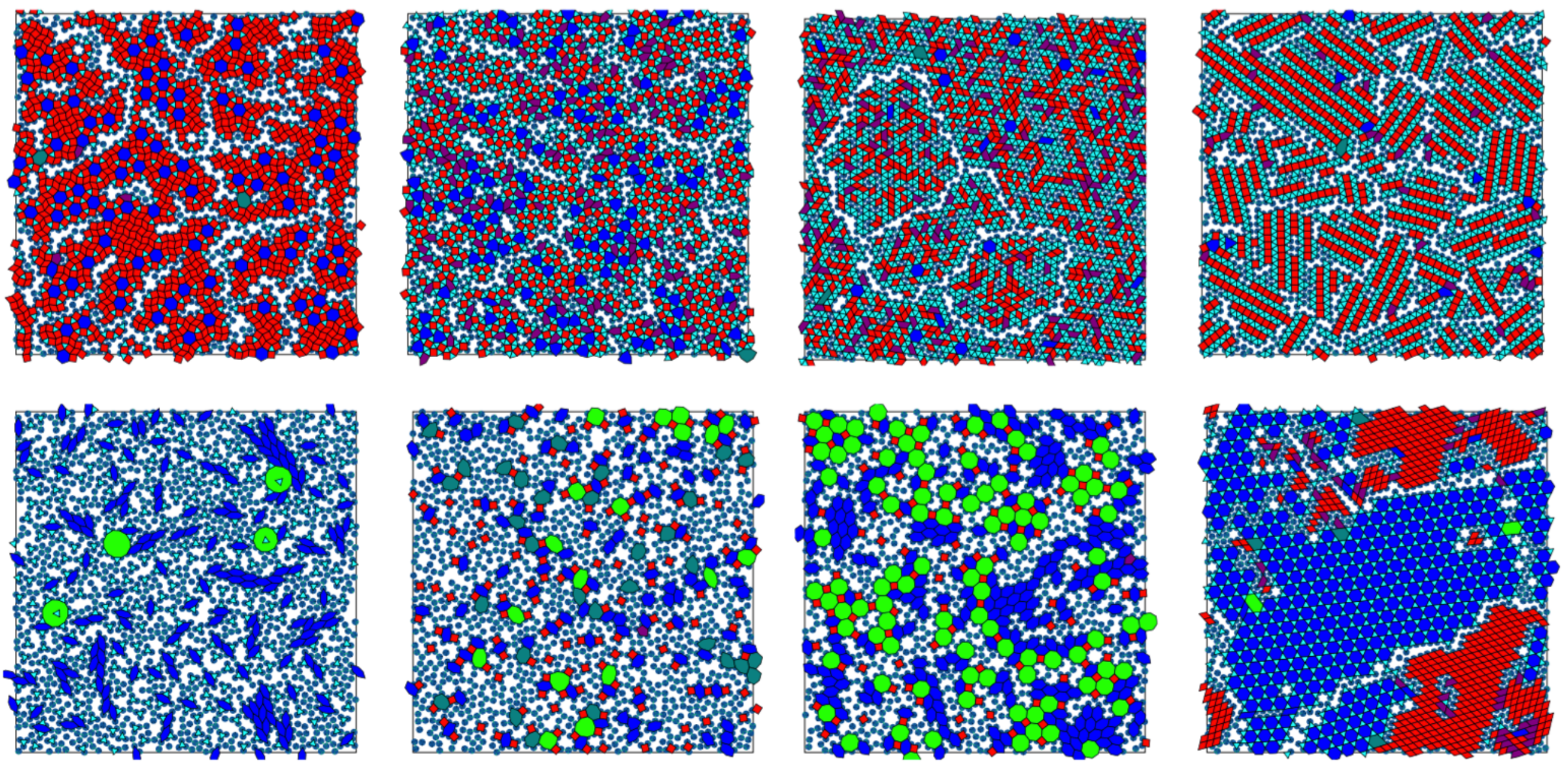} 
   \caption{\label{figs3} The result of simulations done with all patch-patch interactions attractive (panel order is as \f{figs1}). Networks equivalent to the desired tilings do not, in general, assemble.}
\end{figure*}

\break

\begin{figure*}[h!] 
   \centering
\includegraphics[width=0.7\linewidth]{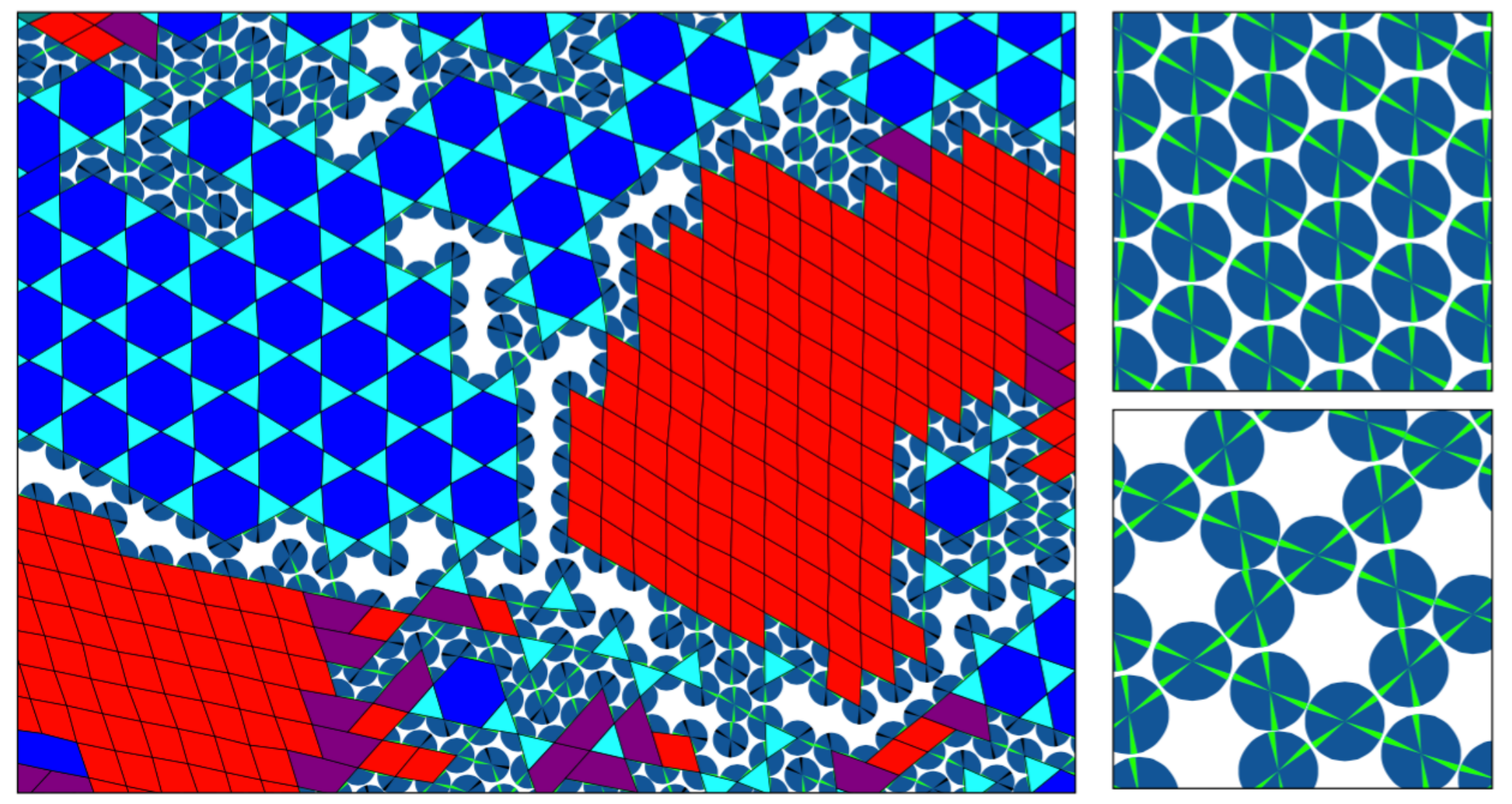} 
   \caption{\label{figs4} The $3.6.3.6$ particle with all patch-patch interactions attractive spontaneously forms two networks (close-ups of which are shown at right). One is the network equivalent to the $3.6.3.6$ Archimedean tiling; the other is a rhombic network. The latter is denser and so is favored at finite pressure, consistent with the findings of Ref.\c{antlanger2011stability}. With minimal positive design, i.e. using only the interactions shown in Table~\ref{tab1}, only the desired network assembles.}
\end{figure*}

\end{document}